\documentclass{article}
\usepackage{graphicx} 
\usepackage[T1]{fontenc}
\usepackage{tgpagella}
\usepackage[parfill]{parskip} 
\usepackage{array}
\usepackage{graphicx}
\usepackage[onehalfspacing]{setspace}
\usepackage{booktabs} 
\usepackage[numbers]{natbib} 
\usepackage{acronym}
\usepackage{appendix}
\usepackage{float}
\usepackage{etoolbox}
\usepackage{amsmath}
\usepackage{xurl}
\usepackage{wrapfig}
\usepackage{chngcntr}

\title{Friends as Sensors: High-Sensitivity Monitoring of Misinformation and Platform Dynamics via the Friendship Paradox}

\author{
  \begin{minipage}{\dimexpr\textwidth-12pt}
    \centering
    Mathias Angermaier\textsuperscript{1},
    Jana Lasser\textsuperscript{1},
    Stephan Lewandowsky\textsuperscript{2, 3},
    David Garcia\textsuperscript{4} \\[0.5em]
    \small
    \textsuperscript{1}University of Graz \quad
    \textsuperscript{2}University of Bristol \quad
    \textsuperscript{3}University of Potsdam \quad
    \textsuperscript{4}University of Konstanz
  \end{minipage}
}

\begin{document}


\maketitle

\begin{abstract}
The friends of a randomly chosen social media user are, on average, better connected than the user, a network property known as the friendship paradox. We use it to monitor misinformation on Twitter/X around the 2024 US election, comparing roughly 27 million tweets from random users with 31 million from the accounts they retweet, which we call sensors. Sensors share news of about the same quality as random users, because news quality follows the political orientation of linked outlets and not the user's position in the network. News links rated by NewsGuard reach sensors a median of 70 minutes earlier, across the quality spectrum. Over the election window, right-leaning content rises about 2.5-fold in visibility, measured as impressions per tweet, among sensors but not among random users. Friendship-paradox sampling therefore cannot separate low-quality from high-quality news, but it detects news early and tracks what a platform makes visible.
\end{abstract}

\section{Introduction}

Detecting misinformation early in its spread on social media is a recurring challenge for empirical research and platform governance in the attempt to safeguard public discourse.
While misinformation is only a small fraction of overall media consumption \cite{allen2020evaluating}, its potent spread \cite{vosoughi2018spread} raises the question of how it can be detected and counteracted before reaching wide audiences.

Existing interventions such as accuracy nudges \cite{pennycook2021accuracy}, inoculation \cite{vanderlinden2017inoculating, roozenbeek2019inoculation}, and corrections \cite{bode2015corrections} can demonstrably reduce belief in misinformation and its sharing in experimental settings. However, these approaches typically intervene after misinformation has been identified, instead of at the start of a sharing cascade. Corrections face the additional challenge that misinformation continues to influence judgments even after it has been debunked \cite{lewandowsky2012misinformation}, which makes early intervention all the more important. Bak-Coleman et al. \cite{bakcoleman2022combining} showed this by simulating intervention strategies on 10.5 million tweets from misinformation events around the 2020 US presidential election. In their model, removing a false claim within 30 minutes cuts its total spread by about 94 percent. The same removal after four hours cuts it by only 56 percent, and no single intervention was sufficient on its own at realistic levels, so countermeasures have to come early.

\begin{wrapfigure}[18]{r}{0.6\textwidth}
\vspace{-0.6cm}
\centering
\includegraphics[width=0.5\textwidth]{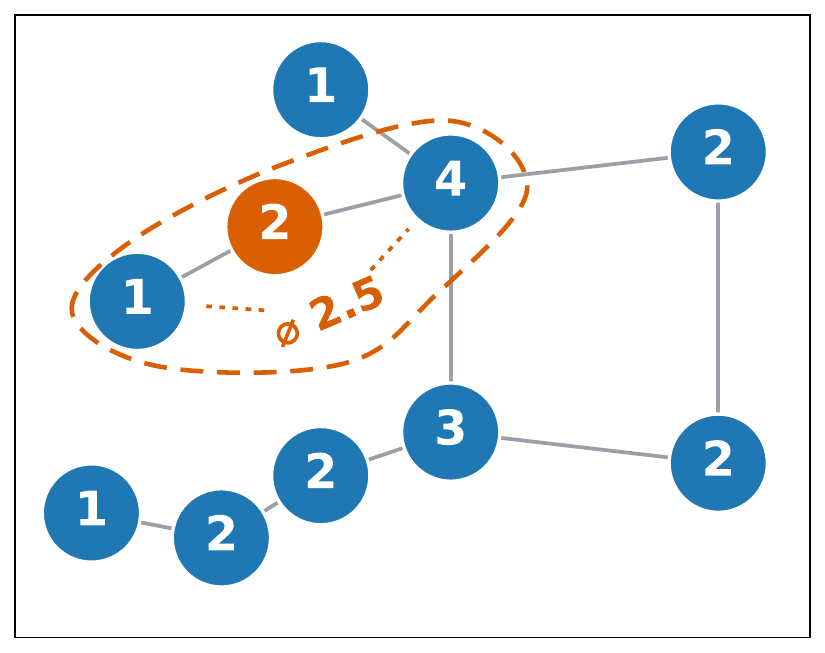}
\caption{The friendship paradox in a network of ten people, who average 2.00 friends while their friends average 2.47. In the circled group the orange node has 2 friends, these friends have 1 and 4 friends, so the friends average 2.5, more than the node's own 2.}
\label{fig:fp_illustration}
\end{wrapfigure}

We use the friendship paradox \cite{feld1991original}, a network property according to which an individual's friends have, on average, more friends than that individual, to identify users who are positioned to receive content earlier than randomly sampled users.
Those more centrally positioned users are referred to as sensor users.
This counterintuitive phenomenon arises from a fundamental sampling bias: individuals with many connections (high-degree nodes) appear as friends to numerous people, while less-connected individuals appear to fewer. As a result, the popular individuals are counted many times when we look at the friends of others, and they pull the average up. Figure~\ref{fig:fp_illustration} shows a small example network that illustrates this mechanism. On Twitter/X the paradox is especially pronounced. Upwards of 90\% of users have fewer followers than the accounts they follow \cite{alipourfard2020friendship, Hodas_2013}.

In previous studies, the friendship paradox has enabled early detection of contagious outbreaks \cite{christakis2010social} and content diffusion on social media \cite{Garcia_Herranz_2014}. A study about a flu outbreak on a campus detected epidemic progression 13.9 days earlier in the friend group than the general population \cite{christakis2010social}. More comparable to our study, García-Herranz et al. \cite{Garcia_Herranz_2014} applied the friendship paradox to Twitter/X by constructing a sensor group from the friends of randomly selected users and tracking hashtag adoption in both samples in parallel. Because high-degree users are overrepresented in the sensor group, hashtags trending in the broader network appeared there first on average, and about 7 days earlier than in the random sample. We extend this methodology to misinformation, using two large-scale Twitter/X datasets (approximately 27 million tweets from randomly selected users and 31 million from their network neighbors) to compare information quality and temporal dynamics, as well as time advantage and exposure of (mis)information between these two samples. Our study is guided by three research questions:

\textbf{RQ1:} Do users occupying more central network positions, as identified through the friendship paradox, share more misinformation than randomly selected users? 

\textbf{RQ2:} Can the friendship paradox be leveraged to detect misinformation sharing earlier than would be possible through random observation of the network?

\textbf{RQ3 (exploratory):} Does friendship-paradox sampling reveal platform-level shifts in content exposure that remain invisible in random samples?

Unlike RQ1 and RQ2, the third question was not part of the original study design. It emerged when the temporal analyses revealed asymmetric impression dynamics in the sensor sample, and we report the corresponding results as exploratory findings.

To foreshadow briefly, the answer to RQ1 is no. Sensor users do not share more misinformation than randomly selected users, and the divide in information quality runs along political orientation, not network position. The answer to RQ2 is yes, but without a quality bias. News links propagate through sensor users with a median lead of 70 minutes over random users, and this lead applies equally to low- and high-credibility sources. The most consequential findings concern RQ3, because over the examination period from June to October 2024 impressions per tweet on right-leaning content in the sensor sample rise roughly 2.5-fold while left-leaning impressions remain flat, and the same asymmetry persists at the user level when posting activity is weighted by users' sharing orientation. None of this is visible in the random sample, where the right-leaning series stays flat at the content level and even drifts slightly downward at the user level. We therefore argue that friendship-paradox sampling is a tool for monitoring platform-level exposure dynamics and a poor detector of content quality.

\section{Results}

\subsection{Descriptive Dataset Analysis}\label{sec:desc_data_analysis}

Table \ref{tab:descriptive} summarizes dataset characteristics. The stats for the random dataset are shown once for the whole dataset and once for the subset that has a time overlap with the sensor data (June-October 2024) as this subset will be used in the comparative temporal analysis. 

\begin{table}[ht]
\centering
\caption{Descriptive statistics for random and sensor datasets.}
\label{tab:descriptive}
\renewcommand{\arraystretch}{1.2}
\resizebox{\textwidth}{!}{%
\begin{tabular}{lrr>{\raggedleft\arraybackslash}p{3.6cm}}
\toprule
\textbf{Metric} 
& \textbf{Sensor} 
& \textbf{Random}  
& \parbox[c]{3.6cm}{\centering\textbf{Random -- Sensor \\ period overlap}} \\
\midrule
Messages (tweets) & 30,964,074 & 27,395,048 & 4,940,620 \\
Unique authors & 9,121 & 9,741 & 9,719 \\
Retweet share & 40.5\% & 47.2\% & 53.8\% \\
Tweets with NewsGuard score & 3.49\% & 0.81\% & 0.37\% \\
\midrule
\textit{Per-author statistics} & & & \\
Messages mean (median) & 3,395 (911) & 2,812 (716) & 508 (163) \\
Messages Gini & 0.747 & 0.746 & 0.714 \\
Impressions mean (median) & 56.7M (3.8M) & 325K (20K) & 104K (4,934) \\
Impressions Gini & 0.897 & 0.910 & 0.926 \\
NewsGuard score mean & 82.88 & 80.37 & 80.77 \\
\bottomrule
\end{tabular}}
\end{table}

Three key observations emerge from these analyses. The sensor sample's mean impressions are roughly $175 \times $ higher than those of the random sample (56.7M vs. 325K), confirming their more central network positions via the friendship paradox. Impressions are the metric to show how many times a post has been seen by other users.
Both samples exhibit high inequality in tweet production and reach (Gini > 0.74), and the sensor sample contains roughly four times more NewsGuard-rated content per tweet (3.49\% vs. 0.81\%), reflecting more active news sharing among central users.

\subsection{Information Quality: Overall Comparison}

\begin{figure}[ht]
\centering
\includegraphics[width=\columnwidth, height=0.5\textheight, keepaspectratio]{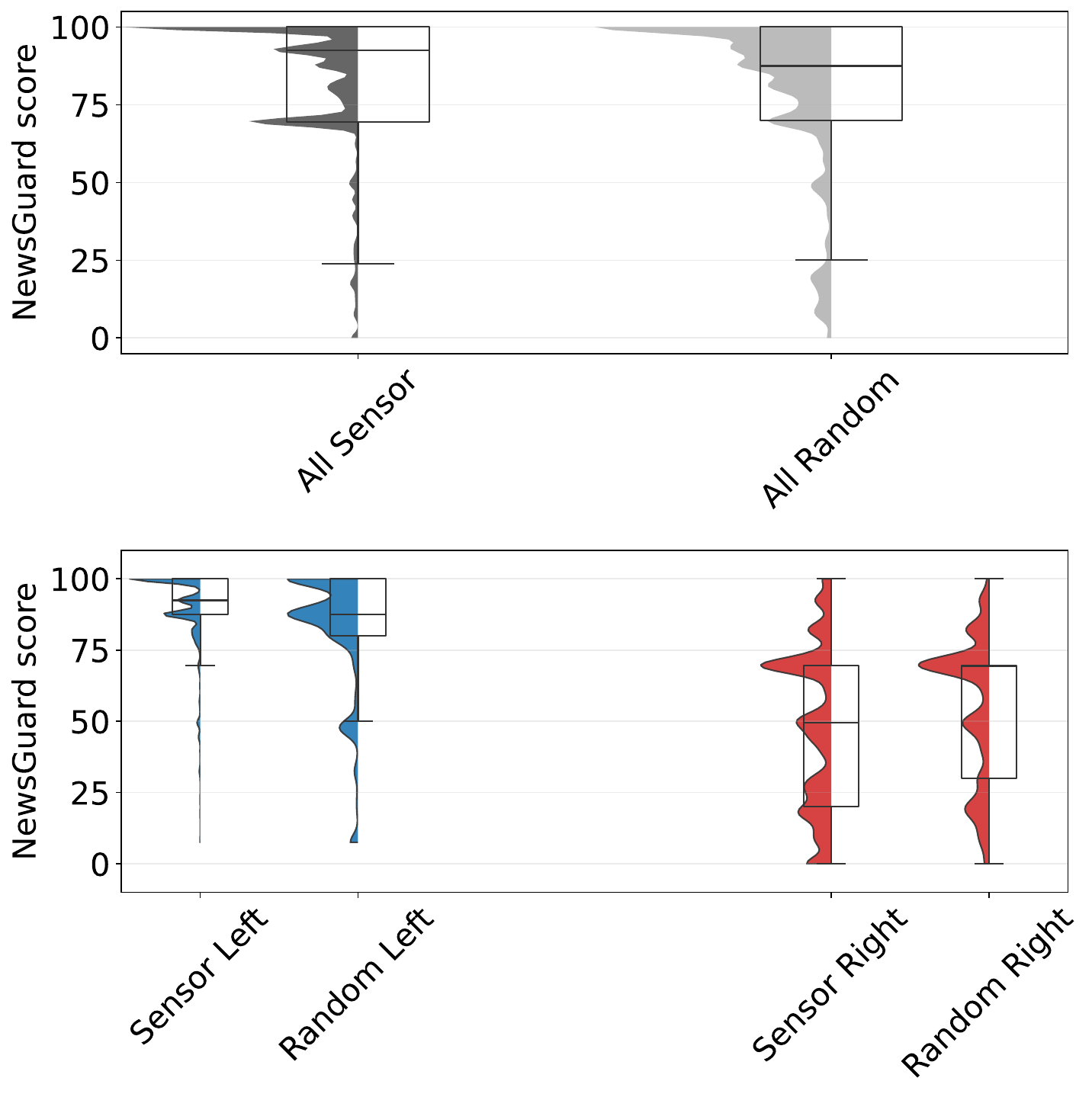}
\caption{NewsGuard score distributions for sensor and random datasets. Overall distributions show no perceptible difference in information quality between the two groups (sensor Q1--Q3: 69.50--100, Random Q1--Q3: 70.00--100). When split by political orientation, however, a clear ideological divide emerges: left-leaning outlets score consistently high (sensor Q1--Q3: 87.50--100, Random Q1--Q3: 80.00--100), while right-leaning outlets score substantially lower, regardless of network position.}
\label{fig:ng_combined}
\end{figure}

Figure \ref{fig:ng_combined} shows NewsGuard score distributions for sensor and random datasets. We find no meaningful difference in overall information quality between high-degree (sensor) and random users when examining raw distributions. The generalized friendship paradox would predict such a difference only if information quality rose with a user's connectivity, and our data suggest it does not. Both distributions are broadly similar, with comparable shapes and central tendencies. The co-occurrent peaks in both datasets arise because NewsGuard scores are not continuous. Each score is the sum of nine binary yes or no criteria with different point values, ranging from 22 points for not repeatedly publishing false or egregiously misleading content down to 5 points for naming content creators, so only a limited set of score values can occur.

A formal statistical comparison reveals a small but statistically significant difference in NewsGuard scores between the two samples, with a median score of 90 in the random dataset and 91.25 in the sensor dataset. The estimated median difference is $\Delta = 1.232$, with a 95\% confidence interval of $[0.724, 1.566]$. The corresponding effect size is Cliff’s $\delta = -0.056$, and a Mann–Whitney $U$ test yields a $p$-value of $8.82 \times 10^{-6}$.

The near-zero Cliff's delta indicates that knowing whether a user is part of the sensor or random group provides essentially no information about the quality of the media outlet they share. Whether central users share better or worse information was an open question with arguments on both sides. Friends have been argued to receive higher quality information than random users \cite{Hodas_2013}, although that conclusion rested on popularity as a proxy for quality. The rapid spread of false news raised the opposite concern, namely that highly connected users could act as hubs of misinformation, even though the spreaders of false news in that study were not better connected than other users \cite{vosoughi2018spread}. Our data support neither direction, because information quality is essentially independent of network position. The generalized friendship paradox therefore does not transfer to misinformation operationalized as the NewsGuard score of shared news domains.

At first glance, the higher reach of sensor users (mean impressions 56.7M vs.\ 325K) hints towards different outreach dynamics, maybe with dynamics that favor a certain kind of news quality. An exposure-weighted analysis on the overlap window confirms that sensor users generate substantially higher weighted misinformation intensity (Cliff's $\delta = -0.399$). However, this finding is largely tautological, as it describes how users with more followers have more outreach. Sensor users amplify all content proportionally to their audience size. In fact, among original tweets the NewsGuard score correlates positively with log impressions in both samples ($r = 0.24$ sensor, $r = 0.21$ random, both $p < 0.001$), so higher-quality content receives greater reach among central and randomly sampled users alike. 

\subsection{Information Quality: Political Affiliation}

Figure \ref{fig:ng_combined} also reveals that the divide in information quality runs along the political orientation of the linked outlets and not along network position. Orientation is a domain-level attribute in the NewsGuard data, so we group tweets by whether they link to outlets labeled as left-leaning or right-leaning and compare the NewsGuard score distributions of the two groups within each sample. Tweets linking to left-leaning outlets show consistently high quality, with an interquartile range of 87.50 to 100 in the sensor sample and 80.00 to 100 in the random sample. Tweets linking to right-leaning outlets score substantially lower, with an interquartile range of 20.00 to 69.50 in the sensor sample and 30.00 to 69.50 in the random sample. The pattern holds in both samples, which indicates that the political orientation of the shared outlet, not the network centrality of the user sharing it, predicts the quality of shared news.

This asymmetry sits in the supply of news, not in the users who share it. Low NewsGuard scores concentrate among right-leaning outlets, and the pattern is equally visible in both samples, so being a high-degree sensor user does not amplify the gap. Prior research approaches the same divide from the demand side and shows that fake news exposure and sharing concentrates among conservative users \cite{guess2019prevalence, guess2020exposure, grinberg2019fake}. Osmundsen et al. \cite{osmundsen2021partisan} connect the two sides. People share news that matches their partisan identity, and the low-quality outlets happen to sit mostly on the right. The same sharing motive on both sides therefore produces different amounts of misinformation. None of this depends on a user's position in the network, which fits the gap appearing unchanged across both samples.

\subsection{Temporal Dynamics in Sensor Data}

\begin{figure}[ht]
\centering
\includegraphics[width=\columnwidth, height=0.6\textheight, keepaspectratio]{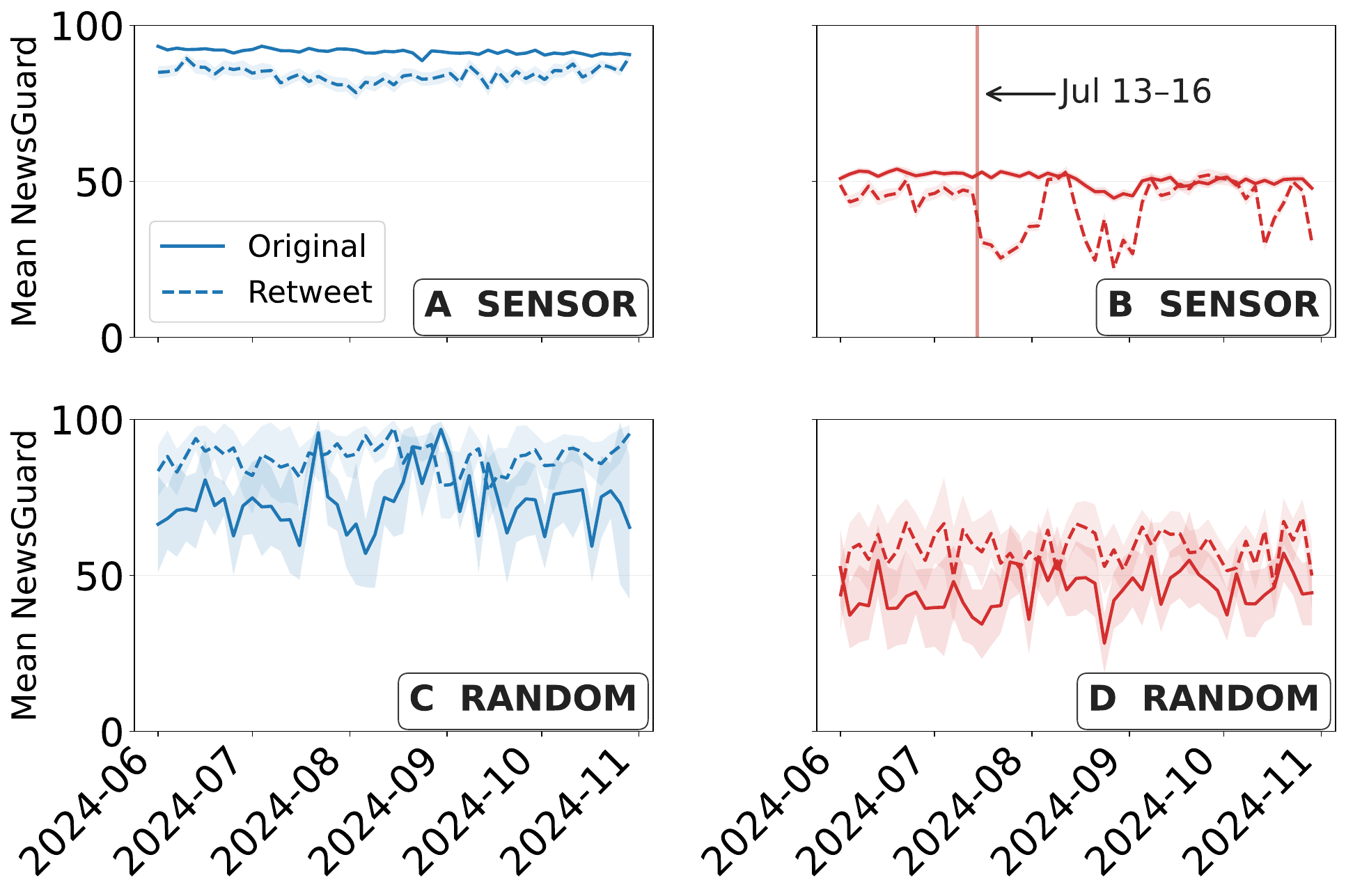}
\caption{Temporal dynamics of NewsGuard scores over 3-day intervals. Tweets are grouped by the political orientation of the news outlets they link to, with original tweets shown as solid lines and retweets as dashed lines. Scores of tweets linking to left-leaning outlets remain stable in both the sensor sample (A) and the random sample (C). In the sensor sample, retweets linking to right-leaning outlets (B) show a notable quality drop around July 16, coinciding with Elon Musk's endorsement of Donald Trump. The corresponding random series (D) shows no comparable pattern, which suggests that the fluctuation is specific to the highly connected sensor network and not a platform-wide trend.}
\label{fig:time_4panel}
\end{figure}

Analysis of NewsGuard scores over time reveals notable dynamics in the sensor dataset and noisy tendencies in the random dataset (Figure~\ref{fig:time_4panel}).

The quality of retweeted right-leaning news shows clear fluctuations over the observation window. The sharpest drop, on July 13--16, 2024, coincides with Elon Musk's endorsement of Donald Trump around which a working paper reports evidence consistent with a platform-side boost to the view counts of Republican-leaning accounts \cite{graham2024algorithmic}. This is additionally marked in Figure~\ref{fig:time_4panel}B. We note as a caution that the same series contains three comparably large swings for which we could identify no external trigger, so we treat this co-occurrence as rather a suggestion.

Summarized, the retweet series in the sensor sample shows distinct and structured fluctuations, while the corresponding random series in Figure~\ref{fig:time_4panel}D does not. Whatever drives these movements operates on the highly connected part of the network and not on the platform as a whole. Whether algorithmic changes were the driver, as suggested by Graham and Andrejevic \cite{graham2024algorithmic}, is beyond what our data can show.

\subsection{Time Advantage Results}

\begin{figure}[h]
\centering
\includegraphics[width=\textwidth]{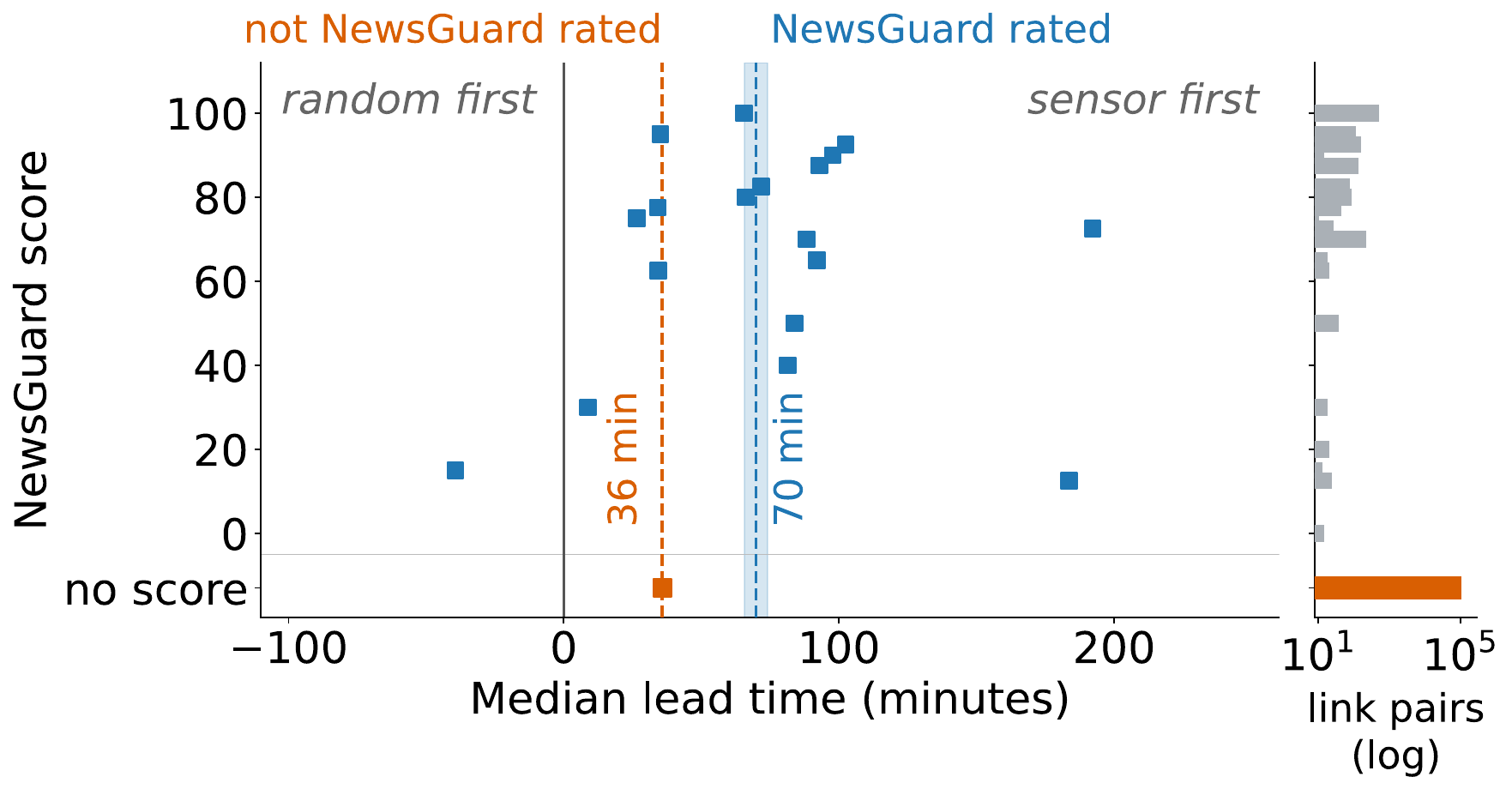}
\caption{Median lead time per NewsGuard score step for links shared in both samples. Each square is a score step, the orange bottom row shows links without a NewsGuard rating, and the right panel gives the number of link pairs behind each row on a logarithmic scale. The vertical line at zero separates links that appeared first in the random sample from links that appeared first in the sensor sample. The orange dashed line marks the median lead of links without a NewsGuard rating (36 minutes), which is indistinguishable from the median across all shared links because such links make up 98.5 percent of all pairs. The blue dashed line and shaded band mark the median lead among NewsGuard-rated links (70 minutes) with its 50\% bootstrap interval. The well-supported steps scatter around the rated-links median across the entire score range, which shows that the lead does not depend on the score ($r = 0.001$, $p = 0.96$ within the well-populated range, $r = 0.051$ across all rated pairs). The uncertainty of individual steps is reported in the text. NewsGuard scores take discrete values, so score levels without any rated outlet in the paired data remain empty. Two thinly supported steps lie outside the plotted range.}
\label{fig:time_advance}
\end{figure}

Comparing first occurrences of links between the datasets (Figure \ref{fig:time_advance}), we find robust evidence that sensor users share news earlier. Among links to NewsGuard-rated news outlets, the median time advantage is 70 minutes (95\% CI 57 to 84) and the mean advantage is 199 minutes (95\% CI 153 to 245). When a rated link is shared in both datasets, it appears first in the sensor sample in 66.9\% of comparisons (95\% CI 65.0 to 68.8). 
Across all shared links, most of which point to other tweets instead of news outlets, the median advantage is 36 minutes (95\% CI 34.6 to 37.4) and the mean advantage is 123 minutes (95\% CI 117.9 to 128.9), with the sensor sample first in 59.9\% of comparisons. News links therefore take longer to cross from the central layer to the random sample than the general link stream, and the sensor lead on news is even more reliable.

Figure~\ref{fig:time_advance} shows that this lead does not depend on content quality. Roughly 88 percent of the rated link pairs are within the score range between 50 and 100. In this span, the time lead is stable and shows no systematic dependence on the score ($r = 0.001$, $p = 0.96$). Across all rated pairs the correlation stays small ($r = 0.051$), reaching significance only through the sample size ($p < 0.001$, $n = 4{,}789$). The 50 percent bootstrap intervals of the individual steps are widest where the pairs panel shows the least support and never shift systematically away from the overall median. Below a score of 50, the estimates rest on few links from few outlets and reflect outlet idiosyncrasies, not quality effects. Some low-scoring outlets reach the sensor sample several hours before the random sample, while others appear in the random sample first. For an early-warning system this picture is encouraging, because the small low-quality outlets that are the natural targets of misinformation monitoring show sensor leads as large as anywhere else.

\subsection{Tweet Exposure}\label{sec:tweet-exposure}

As already assessed in Figures~\ref{fig:ng_combined} and~\ref{fig:time_4panel}, media quality differs along the left- or right-leaning spectrum instead of by the number of connections a user has. Yet the sensor group presents the resulting patterns in a distinctive dynamic manner over time, in contrast to the noisy fluctuations seen in the random group. To assess whether impressions of left- and right-leaning media also evolved differently over the observation period, we performed trend analyses on the geometric-mean impressions computed in 3-day intervals. We applied both an OLS linear regression on the log-transformed means and a non-parametric Mann-Kendall test to avoid relying on linearity assumptions alone.

For right-leaning media, both tests indicated a significant upward trend (OLS: $\beta = 0.019$ log-units per interval, $R^2 = 0.52$, $p < 0.001$; Mann-Kendall: $\tau = 0.54$, $p < 0.001$, Sen's slope $= 0.022$). This corresponds to an approximate 1.9\% increase in impressions per tweet per 3-day interval, compounding to a roughly 2.5-fold increase over the full observation window. The close agreement between the OLS and Sen's slope estimates, along with the relatively high $R^2$, suggests a consistent, structural growth pattern that is not driven by sporadic viral events.

In contrast, left-leaning media showed no significant trend (OLS: $\beta = -0.002$, $R^2 = 0.05$, $p = 0.10$; Mann-Kendall: $\tau = -0.16$, $p = 0.10$), with average impressions per tweet remaining largely stable throughout the period. These results indicate a growing asymmetry in impressions. While left-leaning media maintained a higher baseline of average impressions per tweet at the start of the observation window, right-leaning media average impressions per tweet rose roughly 2.5-fold over the same period, narrowing the gap between right-leaning and left-leaning. Given that the observation period coincides with the 2024 U.S.\ presidential campaign, this finding may in part reflect heightened political mobilization on the right, though disentangling platform-level algorithmic effects from user-driven engagement or paid advertisement campaigns cannot be tested without access to Twitter/X's detailed data on advertisement revenue streams.

\begin{figure}[ht]
\centering
\includegraphics[width=\textwidth]{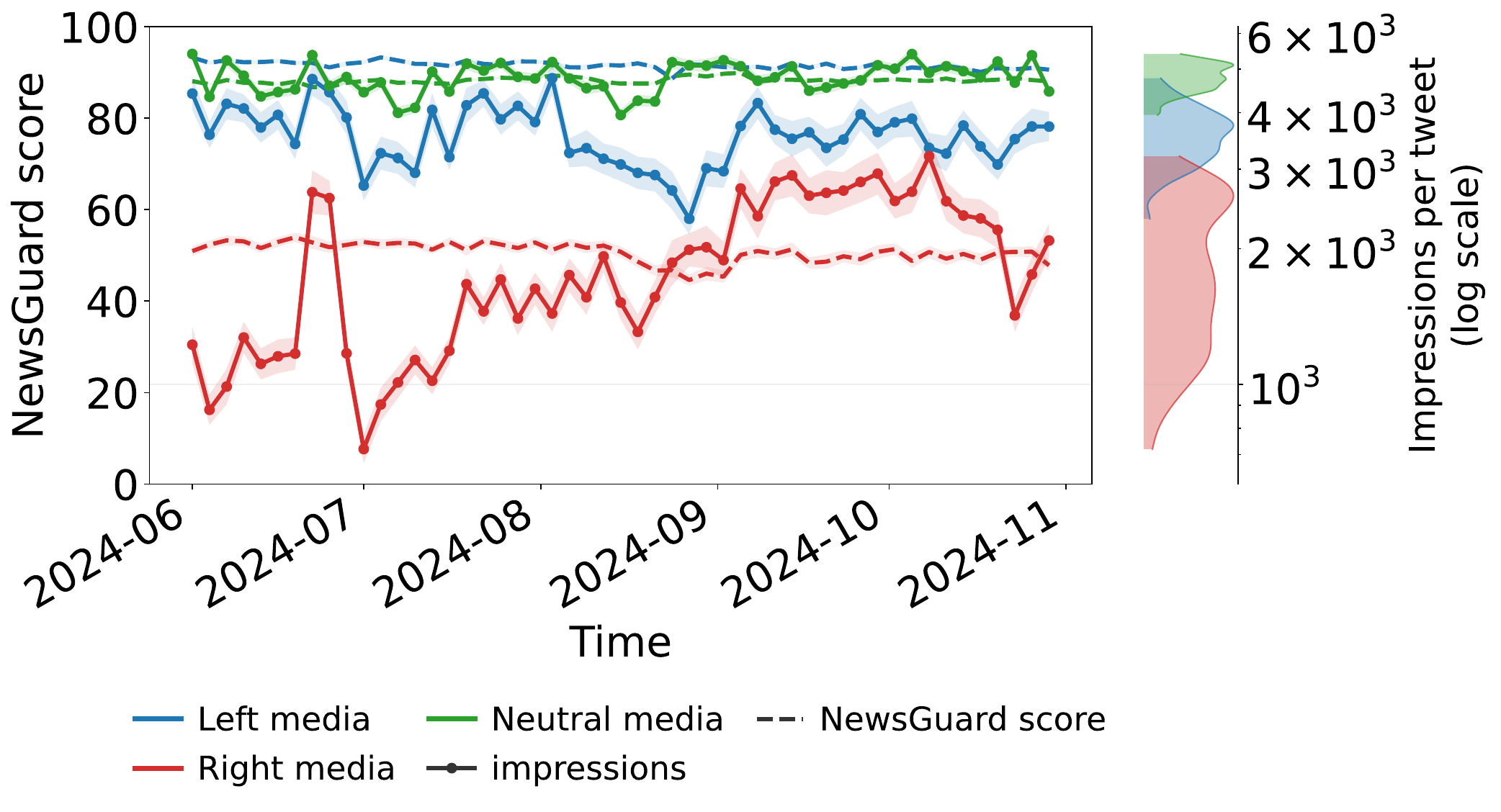}
\caption{Mean impressions per tweet (geometric mean with 95\% CI, log scale) and mean NewsGuard scores for left-, right-, and neutral-leaning media over 3-day intervals. Right-leaning media show a significant upward trend in impressions (OLS: $\beta = 0.019$, $p < 0.001$; Mann-Kendall: $\tau = 0.54$, $p < 0.001$), while left-leaning media remain stable ($p = 0.10$). Neutral-leaning media also show no significant trend (OLS: $\beta = +0.001$, $R^2 = 0.05$, $p = 0.12$; Mann-Kendall: $\tau = +0.13$, $p = 0.18$). NewsGuard scores are read on the left axis and impressions on the right axis. The narrow panel on the far right shows the distribution of the impression values of the three series over the observation window.}
\label{fig:exposure}
\end{figure}

A robustness analysis with different aggregation windows (see Supplementary Note~3) confirms these results. The trend reported so far is measured on tweets that carry NewsGuard-rated links and therefore can be classified by the orientation of the linked outlet. Next, we test whether the same asymmetry is detectable for tweets without NewsGuard rated links and therefore closer to users' \emph{general} posting behaviour. Political tendencies of users are inferred from the author's sharing patterns of NewsGuard rated tweets. Users that do not share classifiable media links are not considered in this analysis. For each sensor user with at least five NewsGuard-rated shares within the observation window (1,815 such users, or roughly 20\% of the 9,121 Sensor authors), we compute three orientation-share indices, $\mathrm{right\_share}$, $\mathrm{left\_share}$, and $\mathrm{neutral\_share}$, which sum to one and give the user's share of NewsGuard-rated tweets that point to right-, left-, and neutral-leaning outlets, respectively. For each of the three orientation fractions, we weight every tweet without a NewsGuard-rated link by the posting user's value of that fraction. Within each 3-day bucket, we then compute share-weighted geometric-mean impressions for three parallel time series. As an example, a tweet with 5{,}000 impressions from a user whose NewsGuard-rated history is 80\% right-leaning, 15\% left-leaning, and 5\% neutral contributes those 5{,}000 impressions to the three bucket-averages with weights 0.8, 0.15, and 0.05, respectively. We refer to the resulting three series as \emph{right-leaning}, \emph{left-leaning}, and \emph{neutral impressions} going forward. Each series therefore isolates, on posts that can not be classified via NewsGuard as they do not contain a media-link, the impression dynamics of users who predominantly share right-, left-, or neutral-rated content.

\begin{figure}[ht]
\centering
\includegraphics[width=\textwidth]{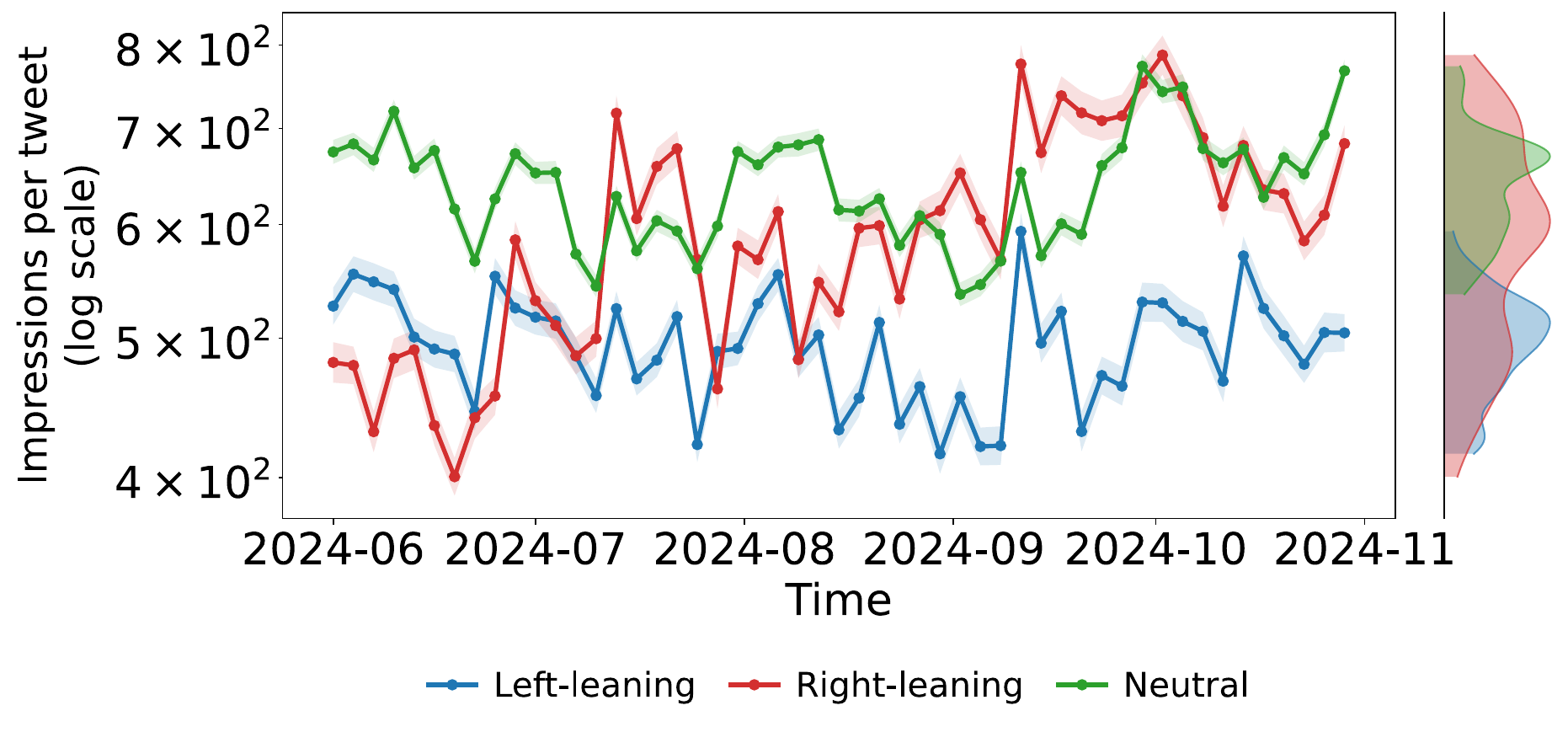}
\caption{Right-, left-, and neutral-leaning impressions on tweets without NewsGuard-rated links (geometric mean with 95\% CI, log scale) over 3-day intervals, by user NewsGuard-orientation share, for the sensor sample. Right-leaning impressions show a significant upward trend (OLS: $\beta = 0.0086$, $p = 6.1 \times 10^{-10}$; Mann-Kendall: $\tau = 0.534$, $p = 3.3 \times 10^{-8}$), while left- and neutral-leaning impressions remain stable. The narrow panel on the right shows the distribution of the three series.}
\label{fig:exposure_weighted_sensor}
\end{figure}

For sensor users, the right-leaning impressions show a significant upward trend that mirrors the trend observed earlier for tweets containing NewsGuard-rated links (Figure~\ref{fig:exposure}) (OLS: $\beta = 0.0086$ log-units per interval, $R^2 = 0.55$, $p = 6.1 \times 10^{-10}$; Mann-Kendall: $\tau = 0.534$, $p = 3.3 \times 10^{-8}$, Sen's slope $= 0.0087$). This corresponds to an approximate 0.88\% increase per 3-day interval, compounding to a roughly 1.55-fold increase over the observation window. In contrast, the left-leaning and neutral impressions both remain flat (left: $\beta = -0.0007$, $p = 0.40$; $\tau = -0.108$, $p = 0.27$; neutral: $\beta = 0.0012$, $p = 0.17$; $\tau = 0.095$, $p = 0.33$), as visible in Figure~\ref{fig:exposure_weighted_sensor}. The asymmetry observed for NewsGuard-rated content therefore extends to users' general posting activity, in that users whose NewsGuard-rated shares lean right experience growing impressions on their non-news tweets as well, while their left- and neutral-leaning peers do not.

To probe whether this user-level amplification is a platform-wide effect or specific to the high-degree sensor sample, we replicate the same procedure on the random dataset (full statistics in Supplementary Note~2). Among random users with at least five NewsGuard-rated shares (272 users, 2.8\% of the random sample, compared to 19.9\% in the sensor sample), the right-leaning trend does not replicate, and mean impressions per tweet drift downward (Mann-Kendall $\tau = -0.216$, $p = 0.026$). Neutral-leaning impressions show a clearer downward trend ($\tau = -0.296$, $p = 2.3 \times 10^{-3}$), and the left-leaning series rises weakly ($\tau = +0.208$, $p = 0.032$). The content-level series of the random sample points the same way, as impressions on right-leaning rated tweets stay flat across the window (OLS $\beta = +0.0007$, $p = 0.85$; Mann-Kendall $\tau = -0.010$, $p = 0.92$), although with a median of 26 rated tweets per bucket this series carries little weight on its own. If the upward trend reflected only growing election-cycle interest in right-leaning content, it would also appear here. Its absence narrows down what happened. Attention did not shift toward right-leaning content across the whole platform, because among randomly sampled users it stayed flat at the content level and declined at the user level. Instead, visibility concentrated on the central accounts sharing right-leaning content. Whether that concentration came from algorithmic ranking, from audiences turning to large accounts during the campaign, or from both, our data cannot decide. A working paper reporting view-count boosts for Republican-leaning accounts in this period \cite{graham2024algorithmic} is consistent with the first reading, but both mechanisms leave the same footprint in impression counts.

\section{Discussion}

\textbf{RQ1: Do high-degree nodes share more misinformation?} No. In terms of median content quality, Sensor (high-degree) users actually share slightly higher quality content than random users (median NewsGuard 91.25 vs 90.00), though the effect size is negligible (Cliff's $\delta = -0.056$). However, when we account for reach through exposure-weighted analysis, Sensor users show substantially higher misinformation intensity (Cliff's $\delta = -0.399$).This reflects amplification and not a tendency to share misinformation. High-degree nodes are not more likely to share it, but what they share reaches larger audiences. Information quality divides along political orientation, and network position plays no visible role. Right-leaning media score substantially lower on NewsGuard and show high variability in quality particularly when retweeted, while left-leaning media maintain consistently high quality scores regardless of network centrality.

\textbf{RQ2: Can the friendship paradox enable earlier detection of misinformation?} Yes, but early detection is not linked to low quality information. Sensor users share news links 70 minutes earlier (median) than random users, with the sensor group posting first in 67\% of comparisons. This time advantage shows no systematic dependence on information quality ($r = 0.001$ within the well-populated score range, $r = 0.051$ across all rated pairs). The general link stream consists mostly of references to other tweets and reaches the random sample 36 minutes after the sensors, while news links take 70 minutes, so news travels along a slower path than the everyday traffic of the platform and monitoring gains a longer head start on news.

\textbf{RQ3: Does friendship-paradox sampling reveal platform-level exposure shifts that random samples miss?} Yes. Over the June -- October 2024 window, mean impressions per tweet on right-leaning media in the sensor sample rise by a factor of roughly 2.5 (Mann-Kendall $\tau = 0.54$, $p < 0.001$), while left-leaning impressions remain flat. The same asymmetry persists at the user level when general posting activity on tweets without NewsGuard-rated links is weighted by users' NewsGuard-orientation share ($\tau = 0.534$, $p = 3.3 \times 10^{-8}$). In the random sample the content-level series stays flat and the user-level series drifts downward. The sensor sample therefore picks up an election-period amplification of right-leaning content that is invisible in the random baseline and even reversed at the user level.

Our findings clarify the function of high-degree nodes for the spread of misinformation in social networks. The work of Vosoughi et al. \cite{vosoughi2018spread} established that false news diffuses farther, faster, and more broadly than true news on Twitter.
Subsequent work by Allen et al. \cite{allen2020evaluating} and Grinberg et al. \cite{grinberg2019fake} contextualized these concerns by demonstrating that fake news represents a small share of overall news consumption and is concentrated among a minority of users. Our findings feed the perspective, in that misinformation sharing reflects specific ideological behavior and not a general property of the social network structure.

The absence of a generalized friendship paradox specifically for misinformation is theoretically significant. The generalized friendship paradox extends to numerous attributes that correlate with network centrality such as happiness \cite{bollen2017happiness}, scientific productivity \cite{eom2014generalized}, or socially influenced behavior \cite{Jackson_2019}. Our finding suggests that misinformation does not function as a centrality-correlated trait. Our partisan-stratified results indicate that the dominant axis of variation in source credibility is political orientation, which is not correlated with network position in our sample. This mirrors the individual-level evidence of Osmundsen et al. \cite{osmundsen2021partisan}. People share fake news out of hostility toward the other political side, not because they cannot tell true from false, and the supply of shareable falsehoods happens to lean right. Neither the hostility nor the supply cares about a user's position in the network. A partisan motivation meeting a partisan supply therefore predicts exactly what we observe, a quality divide that follows political orientation and ignores network centrality.

The second contribution of our study is methodological. Friendship-paradox sampling has been validated as an early-detection tool for hashtag adoption \cite{Garcia_Herranz_2014} and epidemic spread \cite{christakis2010social}. Our results indicate that the same sampling logic can be re-purposed into a monitoring instrument for platform dynamics. The triangulation works through the contrast between the two samples. Both are drawn from the same platform, cover the same period, and are measured with the same instruments, but they sit at opposite ends of the visibility hierarchy. A change that affects the platform as a whole moves both series in the same way. A change in how the platform distributes visibility makes the two series diverge. This offers a way to observe platform dynamics from the outside. The approach requires access to behavioral data through the platform's interfaces, in our case including impression counts, but no access to the ranking systems themselves. The limit of the approach is equally clear, since a divergence between the two samples shows that central and ordinary users experience the platform differently, and it flags when that difference changes, but it cannot name the mechanism behind the change.

The temporal patterns in the sensor data hint towards a recognizable influence of social media platform governance on information quality dynamics. The marked quality drop among retweets of right-leaning outlets in the sensor sample following Elon Musk's endorsement of Donald Trump (July 13, 2024) is consistent with platform-level interventions shifting information quality among high-reach users, although our observational data cannot establish the cause. This co-occurrence highlights the interaction between platform governance and misinformation spread. Changes in platform dynamics, algorithmically or in natural user behavior, can register rapidly among network-central users while remaining invisible in randomly sampled users. We show that the friendship paradox is a suitable method for reaching samples enriched with such central actors who serve as sensitive antennas for detecting these platform-level shifts.

The exploratory findings under RQ3 gain significance when connected to how users form beliefs about the opinion climate. People rely on their local neighborhood to judge how common opinions and behaviors are, and because of homophily, the well-documented tendency to connect with similar others \cite{mcpherson2001birds}, this neighborhood is already a politically narrowed slice of the population. The friendship paradox adds a second distortion on top. Within the narrowed neighborhood, the most-connected users receive disproportionate visibility, which can make globally rare views appear common, a pattern known as the majority illusion \cite{lerman2016majority}. Homophily therefore narrows what a user sees, while the friendship paradox skews whom they see most prominently within that narrowed view \cite{Jackson_2019}. Our two samples make this second distortion measurable. The random sample describes what the network actually produces, while the sensor sample approximates what the network shows when viewed through people's friend lists. The difference between the two is not a nuisance but a quantity of interest, because it is precisely the layer at which perception and reality of the opinion climate can drift apart.

Two of our findings give this layer empirical shape. Sensor users reach on average an audience about 175 times larger than random users (Table~\ref{tab:descriptive}), so the voices a typical user encounters are drawn from a small and highly amplified subset of the network. During the 2024 election window this subset shifted. Impressions on right-leaning content rose about 2.5-fold in the sensor sample while they stayed flat in the random sample, so the visible mix of political content changed. This is the precondition for false consensus effects, in which people misjudge how widely an opinion is shared because their visible sample of the network is skewed \cite{lewandowsky2026architecture}. False consensus beliefs in turn are associated with populist attitudes \cite{steiner2025falseconsensus}. Related evidence shows that engagement-based ranking also distorts perceived conversational norms \cite{brady2026redesigning}, which concerns the form of discourse and not the distribution of opinions, but both lines of research point to the same conclusion, namely that what platforms make visible systematically shapes what users believe about others \cite{lorenzspreen2020behavioural}.

Friendship-paradox sampling offers a way to observe this distortion layer directly. A perception gap measure could compare the politically classified content mix a typical user is exposed to, estimated from the sensor sample and weighted by impressions, with the actual production mix in the random sample. Tracked over time, such a measure would show whether platform changes widen the gap between the seen and the actual network, and it could be validated against survey-based false consensus measures \cite{steiner2025falseconsensus}. We consider this a promising direction for future work.

The practical consequence is that targeting by network position alone will not reach the users most responsible for low-quality sharing, because in the timespan of our dataset the quality divide follows political orientation. The friendship paradox still works as intervention infrastructure. The 70-minute median lead on news links shows no detectable dependence on quality, so it can support general-purpose early-warning workflows, for example queuing emerging content for human review or surfacing accuracy prompts before content reaches viral scale. Beyond misinformation, the contrast between sensor and random samples offers a direct way to observe how far the network that users see drifts from the network that exists, which connects platform monitoring to the study of perceived opinion climates \cite{lewandowsky2026architecture}.

Several limitations require consideration in interpreting our findings. 
\textit{Platform specificity.} Our analysis focuses exclusively on Twitter/X during June--October 2024, a period encompassing the final months of the U.S. presidential campaign and substantial platform policy changes under Elon Musk's ownership. Our analysis is observational, so temporal co-occurrences with platform events cannot establish causal relationships. Misinformation dynamics may differ on other platforms with different network structures, algorithmic curation, and user demographics. The 2024 U.S. election context may also limit generalizability to non-electoral periods or non-U.S. contexts.

\textit{Misinformation operationalization.} NewsGuard scores provide a consistent, scalable measure of source quality, but imperfectly capture misinformation. Domain-level ratings cannot assess individual article accuracy, miss misinformation in text-only posts or non-news content, and may not reflect the specific claims users encounter. Right-leaning outlets more often fail NewsGuard's transparency and editorial criteria, which Lühring et al. \cite{luehring2025best} attribute to differences in editorial practice, not to rater bias. NewsGuard's coverage also concentrates on outlets from North America and Western Europe. Several genuine news outlets from Kenya, Nigeria, and India rank among the most shared unrated domains in our data, so news sharing outside the covered regions is systematically missing from our quality analyses.

\textit{Temporal mismatch.} The sensor dataset spans only June--October 2024, while the random dataset extends from 2015--2024. While we account for that by restricting the time advantage analyses to the overlapping period from June to October 2024, the general comparisons may contain misinformation dynamics from the previous nine and a half years. Information quality and sharing dynamics on Twitter/X may have shifted substantially over this period.

\section{Methods}

\subsection{Dataset Construction}

We constructed two Twitter/X datasets to compare the information sharing behavior of randomly selected users with their network neighbors (sensors), following the friendship paradox methodology established by Christakis and Fowler \cite{christakis2010social} and Garcìa-Herranz et al. \cite{Garcia_Herranz_2014}.

\textbf{Random Dataset.} Authors were drawn from the Twitter/X 1\% random sample stream over a two-week period from September 28 to October 11, 2024, as X extended our access to its API for this project. The resulting sample contained 17,290,963 English-language tweets posted by 6,886,160 distinct accounts. 
Because the stream randomly samples tweets and not users, accounts appear in it with a probability that grows with their activity level. To build a meaningful random user sample, we first selected accounts that had at least 100 lifetime tweets, an account age of at least six months, at least 40 followers and 40 followees, and a follower-to-followee ratio within the central 95\% of the distribution.
This left 4,321,163 eligible accounts, from which we drew 10,000 without replacement with probabilities proportional to the inverse of each account's average posting rate. This cancels the activity bias of the stream and yields a sample that is approximately uniform over eligible active English-speaking accounts and not over the tweets they produce. We then retrieved the full timelines of the selected accounts from January 1, 2015 onwards, resulting in a dataset of approximately 27 million tweets from 9,741 randomly selected authors, spanning January 2015 to early January 2025. Supplementary Note~1 reports the sampling procedure and the sample sizes at each step in detail.

\textbf{Sensor Dataset.} As the list of friends of users (i.e. followed accounts) is not visible on X any more, we operationalize information flow over friendship links through retweets in the data. From among the users retweeted by the users in the random dataset timelines, we selected 10,000 user ids at random. We collected their tweet timelines between June 1 and October 31, 2024 (UTC, both included), resulting in a dataset of 30,964,074 tweets from 9,121 users. Some users were lost as their tweets became unavailable between the time when they were retweeted by the random sample and our data collection. The friendship paradox predicts that these sensor users should be positioned more centrally in the network than the random sample, as high-degree nodes appear disproportionately in the retweeted users lists of randomly selected users.
This friend-of-random sampling strategy has been validated in prior work to achieve early-detection of content diffusion on social media platforms when compared to random sampling \cite{Garcia_Herranz_2014}. Kumar et al. \cite{kumar2024friendship} show that the gain of friend sampling over random sampling depends on the structure of the underlying network, which is why we verify the reach of our sensor sample empirically in Section \ref{sec:desc_data_analysis}.

\subsection{Misinformation Operationalization}

Our central credibility metric is the NewsGuard score, a domain-level rating produced by trained journalists on a nine-criterion rubric. NewsGuard scores measure source-level credibility under this rubric, not factual accuracy of individual articles.
NewsGuard's ratings have been shown to be temporally stable \cite{luehring2025best} and consistent with other news domain rating sets \cite{lin2023correspondence}. Source-level trust ratings in general have been found to track professional fact-checker assessments, even when they come from lay crowds \cite{pennycook2019crowdsourced}. We decided to use the domain-level approach for this observational study, because content-level classification faces scalability limits and adversarial manipulation \cite{shu2017fakenews}.

Throughout the paper, we use \textit{low-NG content} and \textit{high-NG content} to refer to NewsGuard-rated score tendencies. \textit{Misinformation} is referred to as the broader phenomenon as discussed in the Introduction. We use a snapshot of the NewsGuard database from September 2024, overlapping with the observation period.

URLs appearing in tweets were expanded to resolve shortened links (e.g., bit.ly, t.co) to their destination domains, which were then matched against NewsGuard's database. The total share of NewsGuard scores per tweet after resolving the shortened links can be found in Table~\ref{tab:descriptive}. Tweets containing URLs to non-news domains or domains not in NewsGuard's database were excluded from quality analyses but retained for overall dataset description. This approach has limitations, as it cannot assess misinformation in non-news content or text-only posts. However, it provides a consistent and reproducible measure across a large number of tweets.

\subsection{Political Orientation Classification}
NewsGuard categorizes news outlets by political orientation on a left-right spectrum based on editorial positioning, audience ideology, and content analysis. We used these domain-level classifications to label tweets containing NewsGuard-rated URLs as left-leaning or right-leaning based on their linked source. This classification enabled us to analyze engagement patterns (likes, retweets, and impressions i.e., number of times a post was seen by users on Twitter/X) with politically oriented content across the network.

\subsection{Statistical Analysis}

\textbf{Information Quality Comparison.} We compare NewsGuard score distributions between sensor and random datasets using both visual inspection (kernel density plots) and formal statistical tests. Given the large sample sizes, we employ effect size measures alongside significance tests. Cliff's delta ($\delta$) provides a non-parametric effect size, computed as the probability that a randomly selected observation from the random dataset exceeds a randomly selected sensor observation minus the probability of the reverse, with values near 0 indicating no effect and values approaching $\pm$1 indicating complete separation. For this comparison we aggregate at the author level. For each author we average the NewsGuard scores of all rated links they shared over the full collection period, and the tests compare these author means between the two samples.

\textbf{Exposure-Weighted Analysis.} Raw NewsGuard distributions treat all tweets equally, but information exposure depends on reach. We calculate exposure-weighted misinformation intensity by weighting each tweet's misinformation contribution (inverted NewsGuard score) by its impressions. This captures how much low-quality content reaches larger audiences through central users.

\textbf{Time Advantage Analysis.} For links appearing in both datasets during the overlapping period (June -- October 2024), we compare the times at which a URL first appeared in the sensor and in the random group and analyze the time difference. Our primary population are links to NewsGuard-rated news outlets. Links without a rating, which are dominated by references to other tweets, serve as a comparison for the general link stream. Our expectation is that a URL will occur sooner in the sensor than the random group. In addition, we have to account for the substantially higher number of observations in the sensor data, resulting in a much higher baseline probability of any given URL to appear in the sensor data set than the random dataset. To control for this asymmetry, we use a downsampling bootstrap. Within randomly selected 7 day windows from June through October 2024, we take all URLs in the random dataset and the same number of URLs, randomly picked, from the sensor dataset. This step is repeated 1,000 times for multiple random 7-day periods as rolling windows within the time period June -- October 2024. From these results, we calculate the median and mean time advantage of information appearing in the sensor group over the random group for link sharing, with 95\% bootstrap confidence intervals. We also calculate the proportion of comparisons where sensor users posted first. To assess whether lead time depends on content quality, we compute the Pearson correlation between NewsGuard score and time advantage of all URLs that appear in both datasets.

\textbf{Non-Parametric Tests.} Given non-normal distributions and large sample sizes, we employ Mann-Whitney-U-tests for group comparisons, supplemented by Cliff's delta effect sizes. Statistical significance is assessed at $\alpha = 0.05$, though we emphasize effect sizes over p-values given that large samples can render trivial differences statistically significant.

\clearpage

\addcontentsline{toc}{chapter}{Bibliography}
\bibliographystyle{unsrtnat}
\bibliography{reference}

\clearpage

\section*{Data availability}
The tweet datasets analysed in this study were collected through the Twitter/X API under the platform's developer terms. The raw data contain tweet texts, user profiles and platform identifiers and can therefore not be shared. We instead publish dehydrated versions of both datasets, which contain the full engagement, timing and NewsGuard information behind every analysis in this paper while all text, profile and link fields are removed and all identifiers are replaced by salted one way hashes. NewsGuard ratings were obtained under a commercial licence and appear in the published data only as per tweet values without the domains they refer to, so the licensed rating table cannot be reconstructed. The dehydrated datasets and the aggregate tables behind Figure 4 are available on Zenodo (DOI 10.5281/zenodo.22895042).

\section*{Code availability}
All analysis code is available at \url{https://github.com/matrasmus/friends-as-sensors}. The repository contains a notebook that reproduces every figure of this paper from the published datasets and the notebook that generates the published datasets from the raw data.

\section*{Acknowledgements}
This project has received funding from the European Research Council (ERC) under the European Union’s Horizon 2020 research and innovation programme under grant agreement No 101020961 (PRODEMINFO). This project was further funded in part by the Austrian Science Fund (FWF, grant DOI 10.55776/P37280), and M.A. acknowledges the FWF funding. For open access purposes, the authors have applied a CC BY public copyright license to any author-accepted manuscript version arising from this submission.

\clearpage

\section*{Author contributions}

S.L. and D.G. conceived the study. M.A. designed the methodology, collected the data, performed all analyses and wrote the manuscript. J.L., S.L. and D.G. advised on the analyses and revised the manuscript. All authors approved the final version.

\section*{Competing interests}
The authors declare no competing interests.

\clearpage
\appendix
\begin{center}
{\Large\bfseries Supplementary Information}
\end{center}
\setcounter{section}{0}
\setcounter{figure}{0}
\setcounter{table}{0}
\renewcommand{\thesection}{Supplementary Note \arabic{section}}
\renewcommand{\thesubsection}{\arabic{section}.\arabic{subsection}}
\renewcommand{\thefigure}{S\arabic{figure}}
\renewcommand{\thetable}{S\arabic{table}}

\section{Construction of the random sample}
\label{app:random-sampling}

This note walks through the construction of the random dataset step by step, from the raw stream data to the 10,000 accounts whose timelines we finally collected. Three stages lead from one end to the other. We first drew a random sample of tweets, then filtered the accounts behind those tweets down to the ones that are usable for our purposes, and in the end we drew a weighted subsample that undoes the activity bias which comes with sampling users through their tweets.

\subsection{Collection}

We listened to the Twitter/X 1\% random sample stream from late September 2024 onwards and stored every tweet with all available fields in JSON format. For this study we took all tweets that passed through the stream in the two weeks from September 28, 2024 00:00:00 GMT to October 11, 2024 23:59:59 GMT (UNIX timestamps 1727481600 to 1728691199, both included), which came to 50,138,766 tweets. Twitter/X labels every tweet with a detected language, and after keeping only the English ones we were left with 17,290,963 tweets posted by 6,886,160 distinct accounts.

For each of these accounts we then requested basic profile information through the Twitter/X API, and we received profiles for 6,437,883 of them, which is 93.5\% of the accounts observed in the stream. The rest could not be retrieved anymore because the accounts had been deleted, suspended, or set to protected somewhere between the tweet and our profile request. All filters below operate on these 6,437,883 profiles.

\subsection{Eligibility filters}

The 1\% stream samples tweets, not users, so an account shows up in it with a probability that grows with how much it posts. A user sample taken straight from the stream is therefore tilted towards the loudest accounts, and on top of that it contains plenty of accounts that are too new, too inactive, or too thinly connected for the sharing and network measures in this study. We applied three inclusion criteria, each one to what the previous one left over.

\paragraph{Activity and account age.} We kept accounts that had posted at least 100 tweets over their lifetime and that had existed for at least six months before the start of the stream sampling window, which means they were created before March 28, 2024. The first condition makes sure a timeline is long enough to say something about an account's sharing behavior, and the second sorts out accounts whose social ties have not had time to settle. Both conditions held for 5,310,348 accounts, or 82.5\% of the retrieved profiles.

\paragraph{Minimum connectivity.} We kept accounts with at least 40 followers and at least 40 followees, because below that an account hangs too loosely in the follower network for friendship-paradox sampling to mean much. This left 4,548,589 accounts, or 70.7\% of the retrieved profiles.

\paragraph{Follower-to-followee ratio.} For every account we computed the ratio of followers to followees and removed the accounts in the top and bottom 2.5\% of the distribution of the logarithm of this ratio. The trimming cuts away the accounts whose connectivity looks nothing like that of an ordinary user. At one end sit celebrity, media, and institutional accounts with huge audiences and hardly any followees, at the other end sit accounts that follow in bulk without being followed back, which is what bots and spam accounts typically do. The accounts that remain have a follower-to-followee ratio between $0.087$ and $19.39$, with a median of $0.79$. This left 4,321,163 eligible accounts, or 67.1\% of the retrieved profiles, and only 1,669 of them were verified accounts.

\subsection{Activity-corrected subsample}

From these 4,321,163 eligible accounts we drew the final 10,000, without replacement and with inclusion probabilities proportional to the inverse of each account's average posting rate. The posting rate is the number of lifetime tweets from the profile divided by the age of the account in days, with the age counted up to the cutoff date of March 28, 2024, so in practice the weight of an account is its age divided by its tweet count.

The idea is that the chance of landing in a random sample of tweets is roughly proportional to how often an account posts, so weighting by the inverse of that rate roughly cancels the head start of the busy accounts, and what comes out is a sample that is approximately uniform over the population of eligible active English-speaking accounts, not over the tweets that population produces. We checked the correction by comparing the distributions of lifetime tweet counts, account ages, and posting rates before and after the weighted draw.

In the last step we retrieved the complete timelines of the 10,000 selected accounts from January 1, 2015 onwards. Of these, 9,741 accounts had at least one tweet in the collected period, and they constitute the random dataset described in Table~1 of the main text.

\subsection{Sample sizes}

Table~\ref{tab:random-sampling} lists how the sample shrank at each step. The percentages refer to the 6,437,883 accounts with retrievable profile information, since that is the pool the filters actually worked on.

\begin{table}[htbp]
\centering
\begin{tabular}{lrr}
\toprule
\textbf{Step} & \textbf{Remaining} & \textbf{\%} \\
\midrule
\multicolumn{3}{l}{\emph{Tweet level}} \\
Tweets in the 1\% sample, Sep 28 -- Oct 11, 2024 & 50,138,766 & \\
Tweets detected as English & 17,290,963 & 34.5 \\
\midrule
\multicolumn{3}{l}{\emph{Account level}} \\
Distinct accounts posting them & 6,886,160 & \\
Profile information retrieved & 6,437,883 & 93.5 \\
$\geq$ 100 tweets and created before Mar 28, 2024 & 5,310,348 & 82.5 \\
$\geq$ 40 followers and $\geq$ 40 followees & 4,548,589 & 70.7 \\
Central 95\% of follower-to-followee ratio & 4,321,163 & 67.1 \\
\midrule
\multicolumn{3}{l}{\emph{Final sample}} \\
Activity-corrected random subsample & 10,000 & \\
Accounts with at least one collected tweet & 9,741 & \\
\bottomrule
\end{tabular}
\caption{Sample sizes at each step of the construction of the random dataset.
Percentages at the tweet level are relative to all tweets in the two-week
window, and percentages at the account level are relative to the 6,437,883
accounts for which profile information could be retrieved. Each filter is
applied cumulatively to the output of the previous one.}
\label{tab:random-sampling}
\end{table}

For a sense of scale, the 4,321,163 eligible accounts have a median of 9,109 lifetime tweets, a median account age of 1,760 days, a median of 366 followers, and a median of 518 followees.

\section{User-level non-NewsGuard analysis: Random sample}\label{app:user-level-random}

We replicate the share-weighted analysis from the Tweet Exposure section of the main text on the random dataset, restricted to the same June -- October 2024 window and the same 3-day buckets as the sensor analysis. Of the 9,741 Random authors, only 272 (2.8\%) have at least five NewsGuard-rated shares within this window, compared to 19.9\% in the sensor sample. Part of this gap reflects the sensor users' higher posting volume, and part reflects their stronger news-sharing engagement. The smaller pool makes the random series noisier, so the trend estimates below carry wider uncertainty than their sensor counterparts. The right-leaning series among these random users drifts downward (OLS: $\beta = -0.0036$, $R^2 = 0.11$, $p = 0.017$; Mann-Kendall: $\tau = -0.216$, $p = 0.026$). The neutral-leaning series falls more clearly ($\beta = -0.0027$, $R^2 = 0.21$, $p = 8.3 \times 10^{-4}$; $\tau = -0.296$, $p = 2.3 \times 10^{-3}$). The left-leaning series rises weakly ($\beta = +0.0048$, $R^2 = 0.15$, $p = 4.8 \times 10^{-3}$; $\tau = +0.208$, $p = 0.032$). The sensor-side upward trend on right-leaning impressions therefore does not appear in the random sample. The trend that does appear runs in the opposite direction.

\begin{figure}[ht]
\centering
\includegraphics[width=\textwidth]{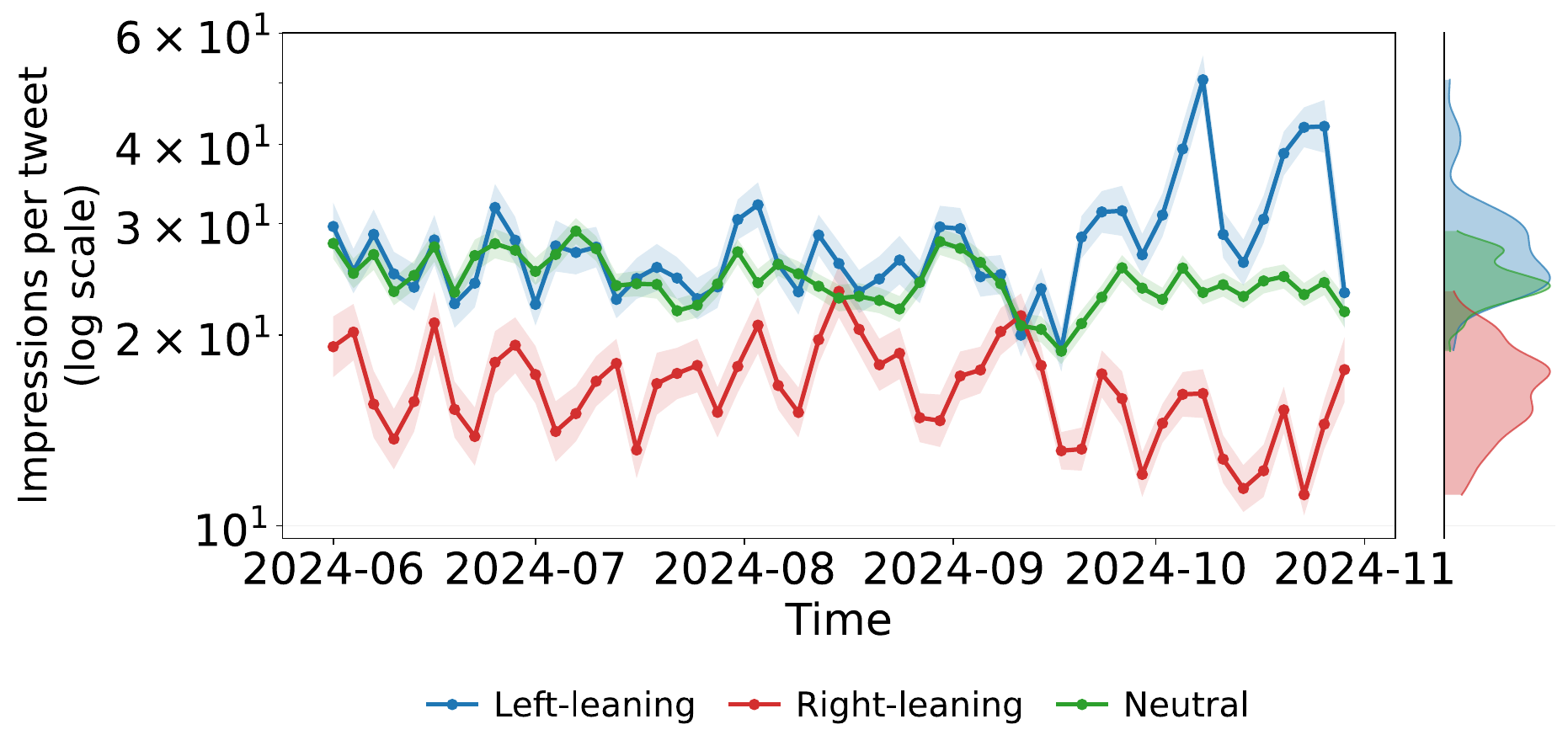}
\caption{Share-weighted average impressions per tweet on tweets without NewsGuard-rated links (geometric mean with 95\% CI, log scale) over 3-day intervals, by user NewsGuard-orientation share, for the random sample. The right-share- and neutral-share-weighted series show small but significant downward trends, while the left-share-weighted series rises weakly. The narrow panel on the right shows the distribution of the three series.}
\label{fig:exposure_weighted_random}
\end{figure}

\section{Bucket-frequency robustness}\label{app:bucket-freq}

One concern with 3-day buckets is that they do not align with the weekly posting cycle on Twitter/X and could therefore be susceptible to day-of-week aliasing. To address this, we re-ran the trend test on the right-leaning original tweets of the sensor sample in the June -- October 2024 window at three aggregation frequencies, namely 1-day (highest resolution), 3-day (used in the main text), and 7-day (immune to day-of-week aliasing because each bucket contains a complete weekly cycle).

Table~\ref{tab:bucket-freq} summarizes the results. The OLS slope scales approximately linearly with bucket length, given that a 3-day bucket spans roughly three times the time of a 1-day bucket and $\beta$ is correspondingly roughly three times larger. The bucket-length-invariant Mann-Kendall $\tau$, in turn, stays in a narrow band of $0.47$ to $0.59$ across all three choices. Our 7-day result confirms the same significant upward trend ($\tau = 0.59$, $p = 9.3 \times 10^{-5}$). The right-leaning impression growth is therefore not an artifact of either daily noise, which would only show up at the 1-day frequency, or weekly-cycle aliasing, which would be eliminated at the 7-day frequency. For the left-only and neutral groups, the no-robust-trend conclusion holds at the 3-day and 7-day frequencies. The marginal nominal significance at the 1-day frequency for left-only ($\tau = -0.119$, $p = 0.030$) and for neutral ($\tau = +0.117$, $p = 0.032$) shows effect sizes roughly four times smaller than the right-only effect at the same frequency, and disappears under any longer aggregation.

\begin{table}[ht]
\centering
\small
\resizebox{\textwidth}{!}{%
\begin{tabular}{lrrrrrrr}
\toprule
Frequency & $n_{\text{buckets}}$ & min tweets/bucket & OLS $\beta$ & $R^2$ & OLS $p$ & MK $\tau$ & MK $p$ \\
\midrule
1-day & 153 & 285     & $+0.0061$ & $0.42$ & $2.1 \times 10^{-19}$ & $+0.472$ & $< 10^{-17}$ \\
3-day & 51  & 1{,}180 & $+0.0187$ & $0.52$ & $2.6 \times 10^{-9}$  & $+0.536$ & $3.0 \times 10^{-8}$ \\
7-day & 23  & 709     & $+0.0421$ & $0.62$ & $8.0 \times 10^{-6}$  & $+0.589$ & $9.3 \times 10^{-5}$ \\
\bottomrule
\end{tabular}}
\caption{Bucket-frequency robustness for the right-only trend on original NewsGuard-rated tweets in the sensor sample.}
\label{tab:bucket-freq}
\end{table}

\end{document}